\documentclass[conference,compsoc]{IEEEtran}
\usepackage{listings}
\usepackage{xcolor}
\usepackage{tikz}
\usetikzlibrary{tikzmark}
\usepackage{booktabs}
\usepackage{multirow}
\usepackage{graphicx}
\usepackage{colortbl} 
\usepackage{lscape}
\usepackage{mathtools}
\definecolor{json_key}{rgb}{0.13, 0.55, 0.13}
\definecolor{json_value}{rgb}{0.0, 0.0, 1.0}
\definecolor{json_string}{rgb}{0.8, 0.0, 0.0}
\definecolor{json_comment}{rgb}{0.0, 0.5, 0.0}

\lstdefinelanguage{json}{
    basicstyle=\ttfamily\small,
    breaklines=true,
    showstringspaces=false,
    captionpos=b,
    frame=single,
    keywordstyle=\color{json_key},
    morestring=[b]",
    morecomment=[s]{/*}{*/},
    morecomment=[l]//,
    morekeywords={true,false,null},
}

\ifCLASSOPTIONcompsoc
  \usepackage[nocompress]{cite}
\else
  \usepackage{cite}
\fi
\ifCLASSINFOpdf
\else
\fi
\usepackage{url}

\begin{document}
%
\title{Explainable Enrichment-Driven GrAph Reasoner (EDGAR) for Large Knowledge Graphs with Applications in Drug Repurposing}




%
\author{\IEEEauthorblockN{Olawumi Olasunkanmi\IEEEauthorrefmark{1},
Evan Morris\IEEEauthorrefmark{2},
Yaphet Kebede\IEEEauthorrefmark{2}, 
Harlin Lee\IEEEauthorrefmark{3},\\
Stanley C. Ahalt\IEEEauthorrefmark{3}, 
Alexander Tropsha\IEEEauthorrefmark{4}, and
Chris Bizon\IEEEauthorrefmark{2}}
\IEEEauthorblockA{\IEEEauthorrefmark{1}Department of Computer Science, \IEEEauthorrefmark{2}Renaissance Computing Institute, \IEEEauthorrefmark{3}School of Data Science and Society}
\IEEEauthorblockA{\IEEEauthorrefmark{4}Division of Chemical Biology and Medicinal Chemistry, UNC Eshelman School of Pharmacy}
University of North Carolina at Chapel Hill, Chapel Hill, NC, USA\\
Emails: \{olawumi, harlin, ahalt, alex\_tropsha\}@unc.edu, \{emorris, kebedey, bizon\}@renci.org}


\maketitle

\begin{abstract}

Knowledge graphs (KGs) represent the connections and relationships between real-world entities. We propose a link prediction framework on KGs named Enrichment-Driven GrAph Reasoner (EDGAR) that infers new edges by mining entity-local rules. This approach is based on enrichment analysis, a well-established statistical method used to calculate mechanisms common to a set of differentially expressed genes. EDGAR's inference results are inherently explainable and rankable, equipped with p-values for statistical significance of each enrichment-based rule.  We demonstrate its effectiveness on a large-scale biomedical KG, ROBOKOP, focusing on drug repurposing for Alzheimer disease (AD) as a case study. Initially, we extracted 14 known drugs from the KG and identified 20 contextual biomarkers through enrichment analysis, shedding light on functional pathways relevant to the shared efficacy of drugs for AD. Subsequently, using the top 1,000 enrichment results, our enrichment-driven system identified 1,246 additional drug candidates for AD treatment. We validated the top 10 candidates using medical literature evidence. EDGAR is deployed within ROBOKOP, along with a web user interface. This is the first work to use enrichment analysis for either large graph completion or drug repurposing. 
\end{abstract}


%
\IEEEpeerreviewmaketitle

\section{Introduction}

Knowledge Graphs (KGs) are powerful frameworks for representing and analyzing complex networks of heterogeneous entities across various domains, such as finance, search engines, social networks, and biomedicine. They integrate diverse data sources into a unified graph structure to reveal hidden patterns and associations under open-world assumptions \cite{shi2018openworld, galkin2022open}. Given the complexity of biomedical research involving various entities (e.g., genes, proteins, chemicals, pathways, diseases), KGs have become essential tools for organizing and exploring such information \cite{fecho2023approach}.

A key challenge in KGs is \textit{edge inference} or \textit{link prediction}, which aims to predict missing relationships between entities based on existing connections. Recent methods for edge inference include embedding-based approaches like TransE \cite{Bordes2013TransE} and RotatE \cite{Sun2019RotatE}, as well as rule-based methods like AMIE+ \cite{galarraga2015fast}, which learn explainable paths. However, rule-based methods face challenges due to the heterogeneity and complexity of KGs. Also, the global nature of the rule limits their adaptability to context-specific relationships, which can lead to false positives.

\begin{figure*}[th]
  \centering
  \includegraphics[width=0.85\linewidth]{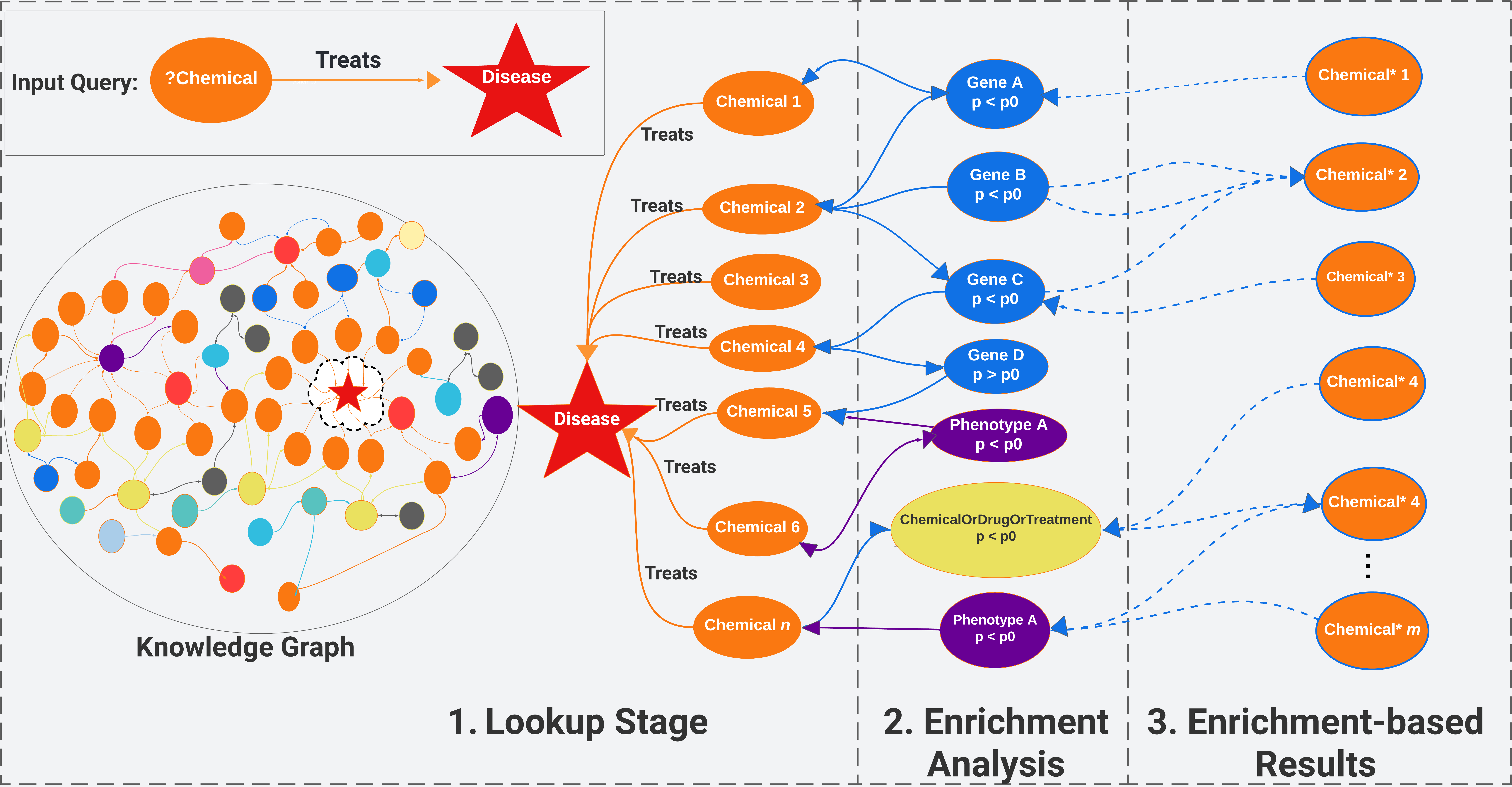}
  \caption{Three stages of EDGAR.
  (1) Subgraph induced by the query includes one-hop neighbors of the disease node (Sec. \ref{sec:lookup_one}).
  (2) Addition of statistically significant enriched nodes and enriched node properties (Sec. \ref{sec:enrich_two}).
  (3) Find one-hop neighbors of enriched nodes and other nodes that share the enriched node properties. These nodes are inferred to have edges with the disease node (Sec. \ref{sec:rec_three}). The p-values from stage 2 propagates to these edges.}
      \label{fig:workflow}
\end{figure*}

In biomedical contexts, edge inference can facilitate \textit{drug repurposing}, where the goal is to identify new therapeutic applications for already-approved drugs.
Drug repurposing benefits from explainable models, as translational researchers are more likely to pursue predictions that can be understood and validated \cite{islam2023molecular}. Research effort \cite{xiao2024repurposing} repurpose non-pharmacological interventions for Alzheimer’s disease. A key challenge is the reliance on literature, which can be biased toward well-studied areas, possibly overlooking emerging or less-documented treatments. \cite{bang2023biomedical} extends the "guilt-by-association" approach to multiple layers of multi-omics data, such as genomics, proteomics, and pharmacology, to strengthen the biological relevance of its repurposing candidates' predictions. However, the complexity of managing multi-layered data can increase computational costs and complicate the interpretation of results, making it more difficult to implement in a clinical setting without robust validation. OREGANO\cite{boudin2023oregano} also explores compounds-disease edge inference through machine learning, strictly by certain types of nodes that are of interest for drug repositioning.

Enrichment analysis (EA) is widely used to interpret biological data by identifying statistically significant associations with functions or pathways. Traditional EA methods, such as Gene Ontology EA \cite{Chenetal2023}, Network-based EA, and Disease EA, employ statistical tests like Hypergeometric or Fisher Exact Test to link genes to biological entities. While traditional EA focuses on gene sets and pathways, we extend enrichment analysis across different entity types using a harmonized KG called ROBOKOP \cite{Bizonetal19}. This approach enables the exploration of drug candidates with similar pathway effects or targets.

To address the aforementioned challenges, we propose EDGAR, a system leveraging local graph enrichment analysis to mine context-specific rules. EDGAR predicts new drug candidates, prioritizes them for further study, and provides interpretable reasoning. For a given disease, EDGAR identifies statistically significant patterns in drugs known to treat that disease and formulates local rules. These rules are then applied to discover new chemicals that share similar characteristics, making them promising candidates for repurposing.

Our contributions are:

\begin{itemize}
    \item A novel enrichment-driven approach, EDGAR, for explainable and rankable edge inference in large, heterogeneous KGs.
    \item A case study on drug repurposing for Alzheimer’s Disease using a biomedical KG, ROBOKOP \cite{Bizonetal19}, marking the first application of enrichment analysis to link prediction and drug repurposing.
    \item Deployment of EDGAR with a web-based interface integrated with ROBOKOP, addressing data scalability.
\end{itemize}

We outline the proposed EDGAR system in Section \ref{sec:edgar},  then describe deployment details in Section \ref{sec:deployment}. Section \ref{sec:alzheimers} introduces a practical use case in drug repurposing for Alzheimer's Disease, and finally, we discuss future research directions and conclude in Section \ref{sec:conclusions}. Additional details and experimental results are deferred to \cite{olasunkanmi2024explainable}.

\section{Enrichment-Driven Graph Reasoner} \label{sec:edgar}

The Enrichment-Driven GrAph Reasoner (EDGAR) integrates a network-based enrichment analysis into a three-hop edge inference in a KG, summarized in Figure~\ref{fig:workflow}. Specifically, we first find known answers to the question, second use enrichment to find their commonalities, and third find new nodes that share those commonalities. We describe each component of EDGAR in detail, using predicting drug candidates suitable for repurposing as a running example. 

\subsection{Lookup of One-hop Neighbors} \label{sec:lookup_one}
A KG is defined as $KG= (V, R)$, where $V$ is a set of entities or nodes, and $R$ is a set of relations or edges. Depending on the node type, the nodes may have properties or data associated with them. We consider KG queries in the form of a triple. A query $(H, r, ?T)$ with entity $H \in V$, node type $T \subset V$ and relation \(r\in R\) asks for a set of all relevant entities $\{t_1, t_2 \ldots, t_n\}$ such that each \(t_i\in T\) satisfies $t_i \xrightarrow{r} H$. 
For example, a potential query for biomedical KGs is ``which drugs treat Alzheimer's Disease (AD)?":
\begin{equation}
   \emph{(H: \textcolor{red}{Alzheimer's Disease})} \xleftarrow{r: treats} \emph{(?T: \textcolor[HTML]{3166FF}{Drug})}. \label{eq:query}
\end{equation}
Note that $T$ is a node type (Drug) while $H$ is a specific node (AD). This query should return a set of $n$ drugs, $L:=\{t_1, t_2 \ldots, t_n\}$, based on the information encoded in the KG: 
\begin{equation}
    \emph{(H: \textcolor{red}{Alzheimer's Disease})} \xleftarrow{r: treats} \{  \emph{\textcolor[HTML]{3166FF}{$t_1, \ldots, t_n$}}\}.
\end{equation}

Responses to this query represent the already known answers to the desired inference query, e.g. drugs that are already annotated as treating AD.  While this is the most generic method to generating the initial set of entities, it can easily be relaxed to begin with, for instance, drugs that are annotated as being investigational or in clinical trials, or for which there may be only limited evidence of efficacy in the literature.

\subsection{Enrichment Analysis of Two-hop Neighbors}\label{sec:enrich_two}
Having found a set of relevant nodes, we now use enrichment analysis methods to discover edges and node properties shared by these nodes more frequently than expected by chance.  

\paragraph{\textbf{Graph enrichment}} Say that $k$ out of $n$ drugs returned from the query $q$ in \eqref{eq:query} are all known to relate to a specific gene $G$. In KG, this connection will be reflected by $k$ edges connecting nodes in $L$ to node $G$:  
\begin{equation}  
  \emph{\{\textcolor[HTML]{3166FF}{$t_1, \ldots, t_k$}\}} ~\overset{r: rel}{\rule{1cm}{0.6pt}} ~\emph{(Differentially expressed gene G)}.
\end{equation}
We purposefully leave $r:rel$ generic, including the direction(s) of the edge.

\paragraph{\textbf{Node property enrichment}} In this scenario, we focus on node properties instead of explicit edge connections. Say drugs \{$t_1, \ldots, t_k$\} share a common node property $d$, such as the binary classification of $t_k$ as a neurotransmitter. 

In both types of lookup results, a natural question arises: “Is gene $G$ or drug property $d$ a key driver in understanding and treating disease $H$, or is this just a coincidence?” To investigate this idea, EDGAR conducts enrichment overrepresentation analysis and calculates statistical significance of the count of enriched node ($G$) or enriched node property ($d$). We emphasize that the enrichment principles extend beyond genes and gene ontology terms to encompass arbitrary node types such as protein, disease, and drug entities, present in the biomedical KG.

The null hypothesis is that the observations are sampled randomly from \textit{hypergeometric distribution}, and there is no biologically meaningful reason behind the fact that $k$ out of $n$ nodes in $L$ are associated with $G$ or $d$. 
The hypergeometric distribution describes the probability of obtaining a specific number of successes in a fixed number of draws without replacement from a finite population containing a known number of successes and failures. Its probability mass function is given by:
\begin{equation}
    P(X=k) = \frac{{K \choose k}{N-K \choose n-k}}{{N\choose n}},
\end{equation}
where:
\begin{itemize}
    \item \(N\) is the number of nodes of type \(T\) in $V$,
    \item \(K\) is the number of nodes of type $T$ in $V$ that relate to $G$ or $d$,
    \item \(n\) is the number of nodes in $L$, and
    \item \(k\) is the number of nodes in $L$ that relate to $G$ or $d$.
\end{itemize}
Since enrichment analyses are interested in \textit{overexpression}, we calculate the survival function, \( \text{SF}(k-1) = P(X \geq k) \), and set that as our p-value. If the p-value is less than a user specified threshold $(p<p_0)$ then we reject the null hypothesis that $G$ or $d$ are associated with the lookup result by chance.  In this case, because our initial set of results preferentially contains a link to $G$ or the presence of property $d$ we treat these as rules that can be exploited to infer new edges.

\subsection{Enrichment-Based Link Prediction between Disease Node and its Three-hop Neighbors} \label{sec:rec_three}

The last stage of EDGAR is using these statistically significant commonalities to infer edges between the disease node and the one-hop neighbors of $G$, or other drug nodes that share property $d$. The p-values from the previous step provide a ranking to these new edges.

We perform a lookup with the updated enrichment rules and return a list of $m$ potential candidate drugs $\{t^*_1, t^*_2 \ldots, t^*_m\} $. We hypothesize that these new drugs could be repurposed to treat Disease $H$. Note that this possibility is not explicitly encoded in the KG, since these nodes were not one-hop neighbors of disease $H$ (i.e. not in initial lookup $L$). EDGAR aims to infer such edges in the KG, ultimately to identify relevant pathways associated with the disease of interest and generate new hypotheses for further research. 

\section{EDGAR in Deployment} \label{sec:deployment}
The EDGAR algorithm is deployed\footnote{\url{https://answercoalesce.renci.org/query}} within ROBOKOP\footnote{\url{https://robokop.renci.org/}, \url{https://github.com/RobokopU24/qgraph}}, an open-source biomedical KG. Here, we describe the ROBOKOP graph and a web user interface for users to visualize and interpret the reasoning results from EDGAR. 

\subsection{ROBOKOP Knowledge Graph}

EDGAR overcomes knowledge-base coverage limitations by leveraging a federated graph,  ROBOKOP \cite{Bizonetal19}, to drive the inference.
ROBOKOP is a heterogeneous biomedical KG that contains approximately 10 million nodes and 250 million edges aggregated from more than 30 knowledge sources. ROBOKOP (Reasoning Over Biomedical Objects linked in Knowledge Oriented Pathways) is an innovative open question-answering platform that leverages numerous openly accessible biomedical databases to explore the interconnections among diverse biomedical data categories. 

Nodes and edges in ROBOKOP follow the ontology and structure in the Biolink model\cite{unni2022Biolink}. These node and edge names must be provided in Compact URI (curie) format that maps to the entity's specific IRI/URI through standardized prefix expansion. For example, the curie representing AD is \textit{``MONDO:0004975"}. ROBOKOP queries and EDGAR-style results are formulated using the NCATS Translator Reasoner API (TRAPI) HTTP API standard \cite{Abuoda}.
Node properties for chemical entities in ROBOKOP are populated from roles defined in ChEBI\cite{hoyt2020extension}, a structured classification of molecular entities of biological interest focusing on `small' chemical compounds. In ROBOKOP, each chemical node, including drugs, has been annotated with its corresponding ChEBI roles, allowing us to categorize them based on their biological or chemical functions, such as antioxidants, antibiotics, and hormones. ChEBI roles turn out to be important enriched node properties in Section \ref{sec:alzheimers}.

\subsection{Web User Interface}
The web UI\footnote{\url{https://edgar.apps.renci.org/}} of EDGAR leverages Dash Plotly \cite{dashplotly} to facilitate seamless interactions. This interface includes a name-to-curie dialogue box, a dashboard that forms a TRAPI query, and a bring-your-own (BYO) data module for visualizing a TRAPI response from EDGAR.

\begin{figure}[http]
  \centering
  \fbox{\includegraphics[trim={0 0 4cm  1.3cm},clip,width=\linewidth]{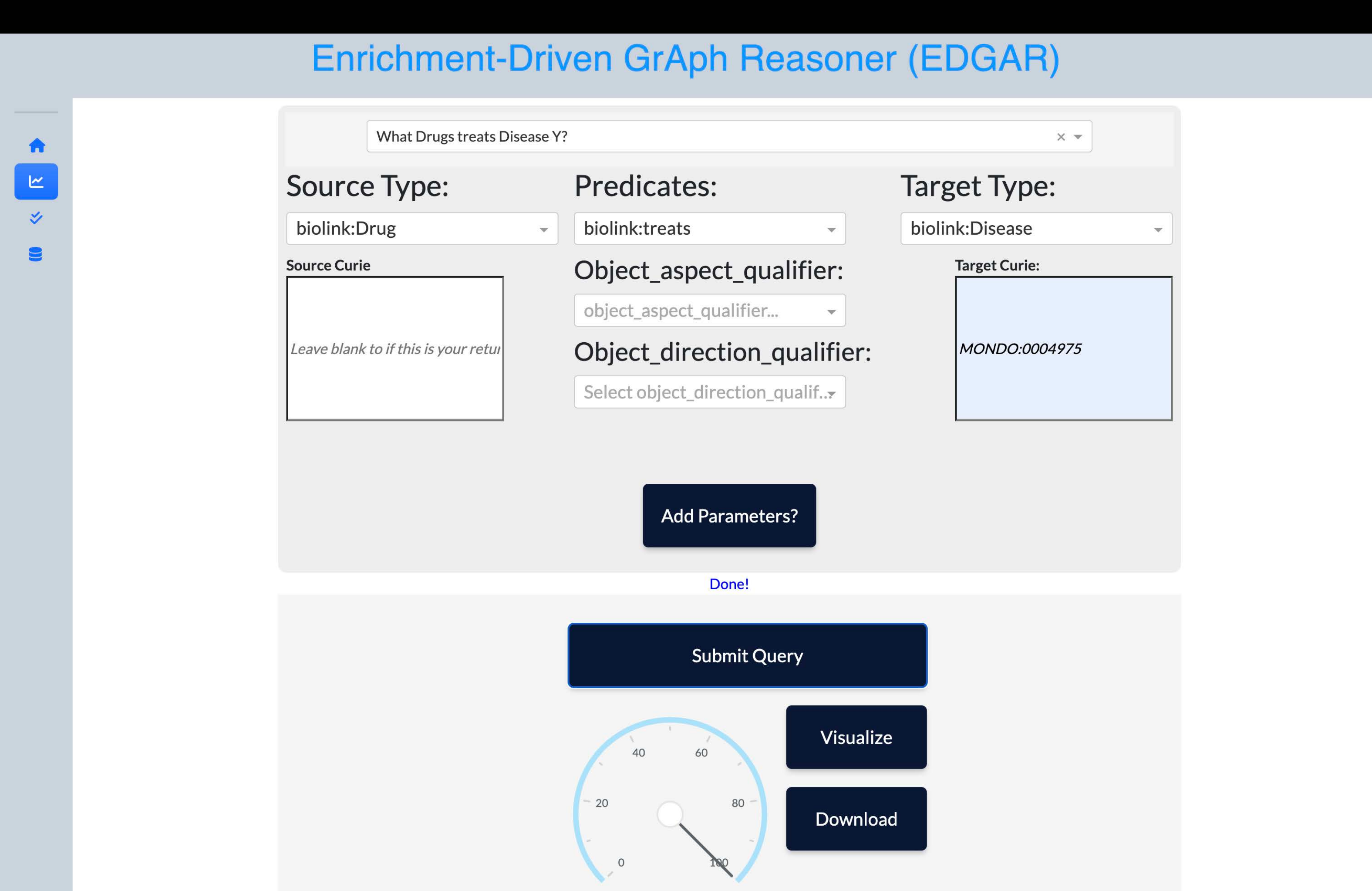}}
  \caption{UI generates valid TRAPI query based on user parameters.}
  \label{fig:ui_trapi}
\end{figure}

The UI page in Figure~\ref{fig:ui_trapi} forms and submits a valid TRAPI query based on user input. To help with the most commonly asked questions, the string literals in Figure~\ref{fig:ui_trapi} feature templates that address:\\ 
\emph{``What Drugs treats Disease Y?"\newline
``What Genes are genetically associated with Disease X?"\newline
``What are the Biological Processes and Molecular Activities that affect Genes X?"}\\
The first template, combined with the curie name for AD found earlier, generated the TRAPI query to be used in Section \ref{sec:alzheimers}. However, we stress that EDGAR can flexibly handle non-templated questions as well. 

\begin{figure}[ht]
  \centering
    \centering
    \fbox{\includegraphics[trim={0 2cm 0  1.3cm},clip,width=\linewidth]{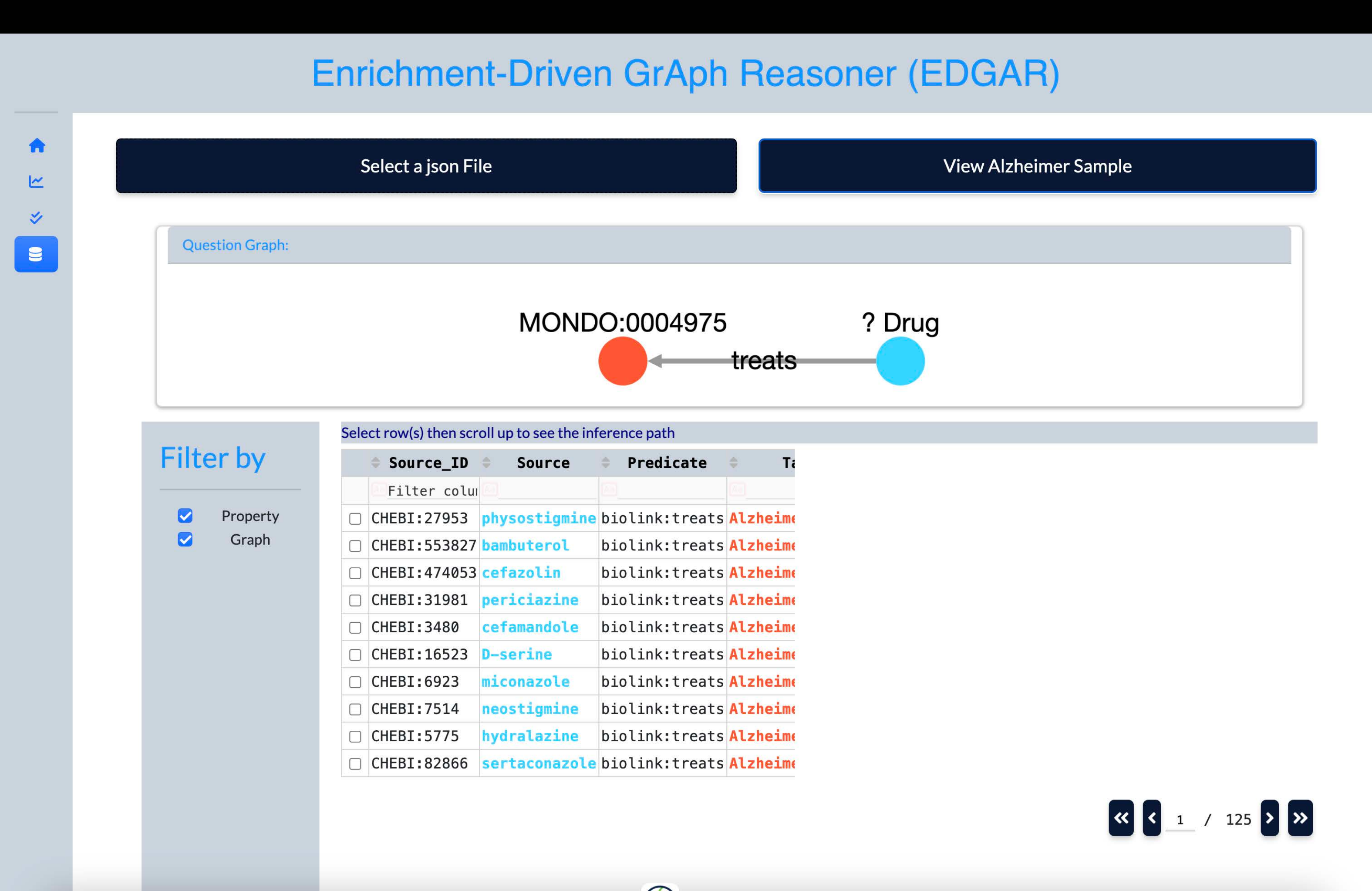}}
    \centering
    \fbox{\includegraphics[trim={0 8cm 0  1cm},clip,width=\linewidth]{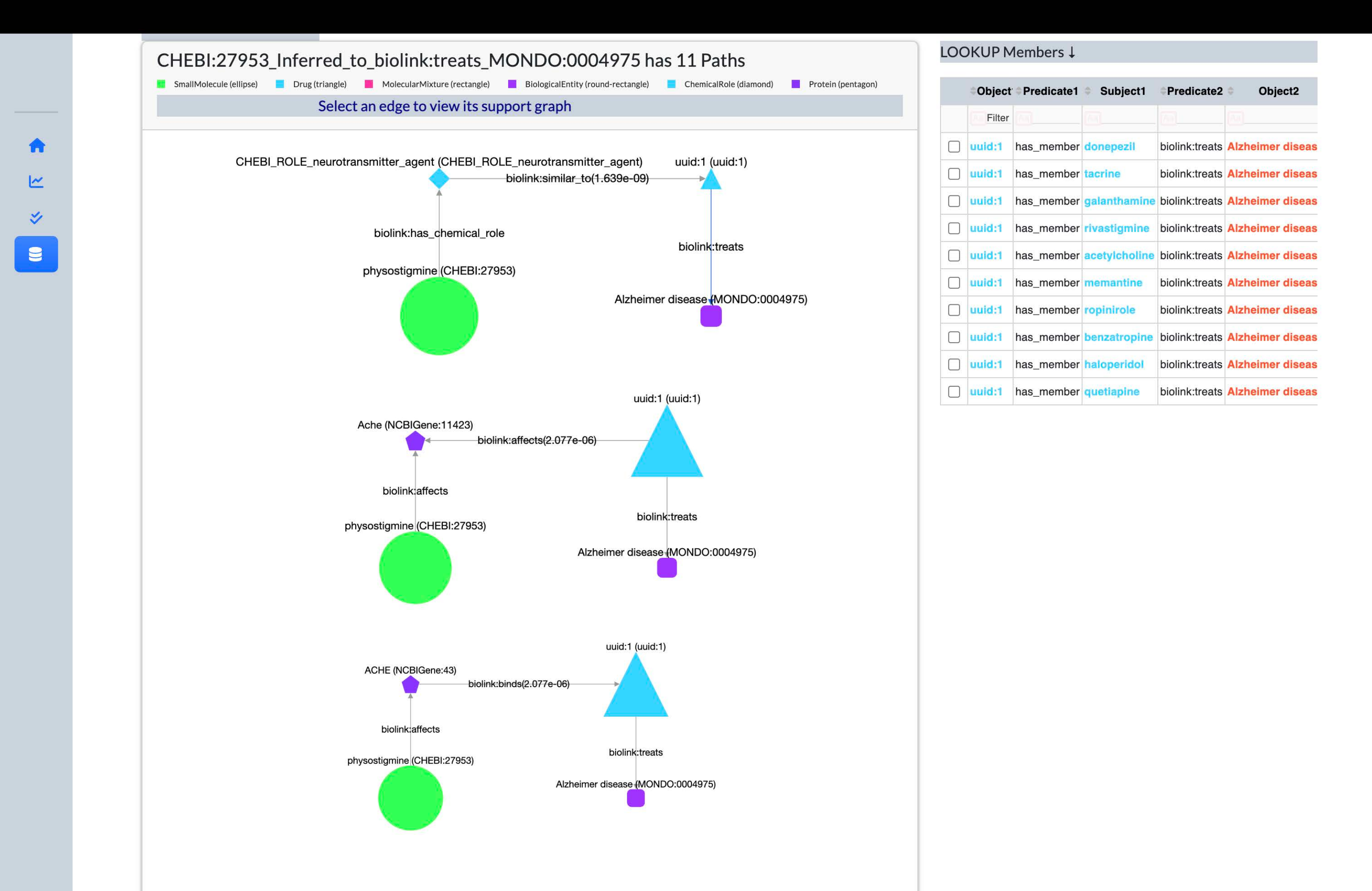}}
  \caption{UI visualizes the enrichment process behind the reasoning.}
  \label{fig:ui_byodata}
\end{figure}

The user can download the response file and upload it to the BYO data dashboard in Figure \ref{fig:ui_byodata} to visualize the outcome of EDGAR. Top of Figure~\ref{fig:ui_byodata} visualizes the query graph (i.e. equation \eqref{eq:query}) and a table of EDGAR result nodes.  
The inferred results are sorted by their p-values in ascending order. When the user clicks on a checkbox, the UI visualizes the enrichment process behind that recommendation (Bottom of Figure~\ref{fig:ui_byodata}) for explainability. For example, the first result Physostigmine (CHEBI:27953) was inferred by 11 different enrichment rules with the top three shown in Figure \ref{fig:ui_byodata}. The first path via CHEBI\_Role\_neurotransmitter\_agent (p-value = 1.63e-9) indicates that though Physostigmine was not initially identified in the one-hop inference, being a neurotransmitter may make it a suitable candidate for the treatment of AD. The top 10 recommendations are discussed and evaluated in the subsequent section.

\section{Case Study: Drug Repurposing for Alzheimer's Disease} \label{sec:alzheimers}

Alzheimer's Disease (AD) is a progressive neurodegenerative disorder characterized by the accumulation of amyloid-\(\beta\) plaques and tau protein tangles in the brain, leading to cognitive decline, memory loss, and behavioral changes \cite{selkoe1994normal, querfurth2010alzheimer}. AD primarily affects older adults, with the incidence rising significantly with age. It is estimated that over 50 million people worldwide are living with dementia, with AD accounting for 60-70\% of these cases \cite{world2019risk}. The aging global population suggests that the number of individuals affected by AD will continue to grow, making it a critical public health challenge \cite{jack2010hypothetical}. Despite significant research efforts, the complex pathophysiology of AD remains only partially understood, with numerous molecular pathways implicated in its progression. As such, AD provides a compelling and urgent case study for the development of advanced computational models, such as knowledge graphs, which can aid in uncovering novel relationships between genes, proteins, and potential therapeutic targets. An enrichment-driven approach to edge inference within such densely connected KG can potentially identify succinct connections and insights that could accelerate the understanding and treatment of AD.

\begin{table}[htp]
\centering
\footnotesize 
\caption{Drugs inferred to have `treats' edge connection with Alzheimer's disease node. Shown are 10 with the lowest p-values.}
\label{tab:alzheimer-treatments-table}
\setlength{\tabcolsep}{2pt}
\begin{tabular}{@{}lll@{}}
\toprule
\multicolumn{2}{c}{\textbf{Drug (Name and curie)}} & \textbf{Enrichment} \\ \midrule
\textcolor[HTML]{000000}{physostigmine} & \textcolor[HTML]{000000}{CHEBI:27953} & property, graph \\ \addlinespace[1mm]
\textcolor[HTML]{000000}{bambuterol} & \textcolor[HTML]{000000}{CHEBI:553827} & property, graph \\\addlinespace[1mm]
\textcolor[HTML]{000000}{simvastatin} & \textcolor[HTML]{000000}{CHEBI:9150} & graph \\\addlinespace[1mm]
\textcolor[HTML]{000000}{quinacrine} & \textcolor[HTML]{000000}{CHEBI:8711} & graph \\\addlinespace[1mm]
\textcolor[HTML]{000000}{ceftibuten} & \textcolor[HTML]{000000}{CHEBI:3510} & graph \\\addlinespace[1mm]
\textcolor[HTML]{000000}{cannabidiol} & \textcolor[HTML]{000000}{CHEBI:69478} & graph \\\addlinespace[1mm]
\textcolor[HTML]{000000}{neostigmine} & \textcolor[HTML]{000000}{CHEBI:7514} & property, graph \\\addlinespace[1mm]
\textcolor[HTML]{000000}{trans-resveratrol} & \textcolor[HTML]{000000}{CHEBI:45713} & graph \\\addlinespace[1mm]
\textcolor[HTML]{000000}{econazole} & \textcolor[HTML]{000000}{CHEBI:4754} & graph \\\addlinespace[1mm]
\textcolor[HTML]{000000}{2-(6R,7R)-7-[[2-(2-amino-4-thiazolyl)-} \\\textcolor[HTML]{000000}{(carboxymethoxyimino)-1-oxoethyl]} \\
\textcolor[HTML]{000000}{amino]-3-ethenyl-8-oxo-5-thia-1-azabicyclo} \\ \textcolor[HTML]{000000}{[4.2.0]oct-2-ene-2-carboxylic acid}& \textcolor[HTML]{000000}{CHEBI:93248} & graph \\
\bottomrule
\end{tabular}
\end{table}

Utilizing enrichment analysis, we identified 1,246 compounds with shared mechanistic pathways that might influence Alzheimer's pathology, i.e. inferred to have `treats` edge connection to AD node in the KG. 10 drugs with the lowest p-values are shown in Table~\ref{tab:alzheimer-treatments-table}.
There were overlapping drugs between the lookup and inferred sets. However, since the lookup drugs are known and used as an input set for enrichment, we are more interested in showcasing the newer drugs inferred by EDGAR. To evaluate, we cross-validate the enrichment-driven drug set with the known treatments and experimental drug in DrugBank and explore additional literature evidence. 

\paragraph{\textbf{Established treatments and mechanistic insights}} Physostigmine and neostigmine are well-established in AD treatment. Physostigmine, a cholinesterase inhibitor, is rapidly absorbed and crosses the blood-brain barrier, effectively treating severe anticholinergic toxicity with central nervous system effects \cite{drugbank}. Similarly, neostigmine is used to reverse the effects of muscle relaxants and treat myasthenia gravis but does not cross the blood-brain barrier, thus limiting its central effects \cite{drugbank}. Both drugs enhance acetylcholine availability in the brain, crucial for alleviating cognitive deficits in AD \cite{bitzinger2016species, malamed2010sedation}.

Despite being a well-established cholinesterase inhibitor for treating AD, Cholinergic dysfunction established in the 80s and 90s might account for why the due was not amongst the initially established treatments identified in the lookup results\cite{cacabelos2020pharmacogenetic}.
   
Simvastatin and trans-resveratrol also stand out as promising candidates for AD repurposing. Simvastatin, an HMG-CoA reductase inhibitor, lowers lipid levels and reduces the risk of cardiovascular events. Recent studies highlight its neuroprotective effects through its impact on inflammation and oxidative stress, critical factors in AD \cite{bansal2023hmg, drugbank}. Trans-resveratrol, a polyphenolic phytoalexin, exists as a stable trans-isomer that resists oxidation under normal conditions \cite{drugbank}. It has shown potential in modulating amyloid-beta aggregation and could play a role in slowing AD progression \cite{cheng2020peripheral, hampel2021amyloid}.
    
Cannabidiol and quinacrine also emerge as intriguing candidates. Cannabidiol, used primarily for managing seizures and neuropathic pain, has demonstrated anti-inflammatory and neuroprotective properties, which could benefit AD patients \cite{drugbank}. Quinacrine, an acridine derivative formerly used as an antimalarial and now as an inhibitor of phospholipase A2, might affect protein aggregation associated with AD \cite{drugbank, salas2021quinacrine}.
 
\paragraph{\textbf{Exploring uncharted territory}}  
    Bambuterol, ceftibuten, and econazole are less directly associated with AD but warrant further investigation. Bambuterol, a bronchodilator, shares commonalities with cholinesterase inhibitors, potentially influencing inflammatory pathways involved in AD \cite{drugbank}. Ceftibuten, a third-generation cephalosporin, and econazole, a broad-spectrum antimycotic, are primarily used to treat infections but could have unexplored impacts on neurodegenerative pathways \cite{drugbank, arumugham2023third}.
    
\paragraph{\textbf{Mechanistic commonality}} The shared mechanisms among these drugs include their effects on inflammatory pathways, oxidative stress, and protein aggregation—central features of Alzheimer's disease. Physostigmine and neostigmine target the cholinergic system, while simvastatin and trans-resveratrol address inflammation and oxidative stress. Cannabidiol and quinacrine contribute to neuroprotection and modulation of protein aggregation. Even though bambuterol, ceftibuten, and econazole are less directly related, their potential impact on AD-related pathways, especially through inflammation and microbial factors, should not be overlooked.

This enrichment-driven analysis not only validates existing therapeutic strategies but also uncovers novel opportunities for drug repurposing, paving the way for innovative approaches in AD treatment.

\section{Conclusion} \label{sec:conclusions}
EDGAR system has a 3 stages process comprising one-hop lookup, enrichment, and enrichment-based link prediction for drug repurposing based on knowledge graph analysis. The one-hop inference identified a set of drugs and enrichment explored their overlapping mechanistic pathways, particularly in complex diseases like Alzheimer's Disease. By focusing on shared mechanistic pathways, this method enhances the accuracy of edge inference, leading to more informed recommendations and the potential identification of novel treatment options. These include both well-established treatments for AD and novel candidates that show promise in modulating key pathological features of the disease. By analyzing these commonalities, we propose an enrichment-driven approach to edge inference in graph-based recommender systems, which can aid in drug repurposing for AD. 

This approach exemplifies the synergy between big data analytics and biomedical research in the context of drug discovery and repurposing. Future research will focus on refining the ranking of new drugs based on their relevance to functional entities, employing latent similarity measures to enhance precision.

\ifCLASSOPTIONcompsoc
  \section*{Acknowledgments}
\else
  \section*{Acknowledgment}
\fi

This work was supported by NIH Grant \#3OT2TR003441-01S2, ARAGORN: Autonomous Relay Agent for Generation of Ranked Networks. The authors would like to thank the ARAGORN team of the Biomedical Translator Project for their valuable contributions.



%
\bibliographystyle{IEEEtran}
\bibliography{main}

\begin{thebibliography}{10}
\providecommand{\url}[1]{#1}
\csname url@samestyle\endcsname
\providecommand{\newblock}{\relax}
\providecommand{\bibinfo}[2]{#2}
\providecommand{\BIBentrySTDinterwordspacing}{\spaceskip=0pt\relax}
\providecommand{\BIBentryALTinterwordstretchfactor}{4}
\providecommand{\BIBentryALTinterwordspacing}{\spaceskip=\fontdimen2\font plus
\BIBentryALTinterwordstretchfactor\fontdimen3\font minus \fontdimen4\font\relax}
\providecommand{\BIBforeignlanguage}[2]{{%
\expandafter\ifx\csname l@#1\endcsname\relax
\typeout{** WARNING: IEEEtran.bst: No hyphenation pattern has been}%
\typeout{** loaded for the language `#1'. Using the pattern for}%
\typeout{** the default language instead.}%
\else
\language=\csname l@#1\endcsname
\fi
#2}}
\providecommand{\BIBdecl}{\relax}
\BIBdecl

\bibitem{shi2018openworld}
S.~Baoxu and W.~Tim, ``Open-world knowledge graph completion,'' in \emph{Proceedings of the Thirty-Second AAAI Conference on Artificial Intelligence (AAAI-18)}, AAAI Press.\hskip 1em plus 0.5em minus 0.4em\relax New Orleans, LA, USA: Association for the Advancement of Artificial Intelligence, 2018, pp. 1957--1964.

\bibitem{galkin2022open}
M.~Galkin, M.~Berrendorf, and C.~T. Hoyt, ``An open challenge for inductive link prediction on knowledge graphs,'' in \emph{Proceedings of the Workshop on Graph Learning Benchmarks (GLB 2022) at The Web Conference 2022 (TheWebConf)}, 2022.

\bibitem{fecho2023approach}
K.~Fecho, C.~Bizon, T.~Issabekova, S.~Moxon, A.~E. Thessen, S.~Abdollahi, S.~E. Baranzini, B.~Belhu, W.~E. Byrd, L.~Chung \emph{et~al.}, ``An approach for collaborative development of a federated biomedical knowledge graph-based question-answering system: Question-of-the-month challenges,'' \emph{Journal of Clinical and Translational Science}, vol.~7, no.~1, p. e214, 2023.

\bibitem{Bordes2013TransE}
A.~Bordes, N.~Usunier, A.~Garcia-Duran, J.~Weston, and O.~Yakhnenko, ``Translating embeddings for modeling multi-relational data,'' in \emph{Advances in Neural Information Processing Systems (NeurIPS)}, 2013, pp. 2787--2795.

\bibitem{Sun2019RotatE}
Z.~Sun, Z.~Deng, J.-Y. Nie, and J.~Tang, ``Rotate: Knowledge graph embedding by relational rotation in complex space,'' in \emph{International Conference on Learning Representations (ICLR)}, 2019.

\bibitem{galarraga2015fast}
L.~Gal{\'a}rraga, C.~Teflioudi, K.~Hose, and F.~M. Suchanek, ``{Fast rule mining in ontological knowledge bases with AMIE+},'' \emph{The VLDB Journal}, vol.~24, no.~6, pp. 707--730, 2015.

\bibitem{islam2023molecular}
M.~K. Islam, D.~Amaya-Ramirez, B.~Maigret, M.-D. Devignes, S.~Aridhi, and M.~Sma{\"\i}l-Tabbone, ``Molecular-evaluated and explainable drug repurposing for covid-19 using ensemble knowledge graph embedding,'' \emph{Scientific Reports}, vol.~13, no.~1, p. 3643, 2023.

\bibitem{xiao2024repurposing}
Y.~Xiao, Y.~Hou, H.~Zhou, G.~Diallo, M.~Fiszman, J.~Wolfson, L.~Zhou, H.~Kilicoglu, Y.~Chen, C.~Su \emph{et~al.}, ``Repurposing non-pharmacological interventions for alzheimer's disease through link prediction on biomedical literature,'' \emph{Scientific reports}, vol.~14, no.~1, p. 8693, 2024.

\bibitem{bang2023biomedical}
D.~Bang, S.~Lim, S.~Lee, and S.~Kim, ``Biomedical knowledge graph learning for drug repurposing by extending guilt-by-association to multiple layers,'' \emph{Nature Communications}, vol.~14, no.~1, p. 3570, 2023.

\bibitem{boudin2023oregano}
M.~Boudin, G.~Diallo, M.~Dranc{\'e}, and F.~Mougin, ``The oregano knowledge graph for computational drug repurposing,'' \emph{Scientific data}, vol.~10, no.~1, p. 871, 2023.

\bibitem{Chenetal2023}
C.~Zhaoyang, C.~Wenjing, L.~Jiao, A.~Zekri, L.~Qing, W.~Haifan, and W.~Xia, ``Gene set enrichment analysis identifies immune subtypes of kidney renal clear cell carcinoma with significantly different molecular and clinical properties,'' \emph{Frontiers in Immunology}, vol.~14, p. 1191365, 2023.

\bibitem{Bizonetal19}
C.~Bizon, S.~Cox, J.~Balhoff, Y.~Kebede, P.~Wang, K.~Morton, K.~Fecho, and A.~Tropsha, ``{ROBOKOP KG and KGB: Integrated Knowledge Graphs from Federated Sources},'' \emph{Journal of Chemical Information and Modeling}, vol.~59, no.~12, pp. 4968--4973, Dec 2019.

\bibitem{olasunkanmi2024explainable}
O.~Olasunkanmi, E.~Morris, Y.~Kebede, H.~Lee, S.~Ahalt, A.~Tropsha, and C.~Bizon, ``Explainable enrichment-driven graph reasoner (edgar) for large knowledge graphs with applications in drug repurposing,'' \emph{arXiv preprint arXiv:2409.18659}, 2024.

\bibitem{unni2022Biolink}
D.~R. Unni, S.~A. Moxon, M.~Bada, M.~Brush, R.~Bruskiewich, J.~H. Caufield, P.~A. Clemons, V.~Dancik, M.~Dumontier, K.~Fecho \emph{et~al.}, ``Biolink model: A universal schema for knowledge graphs in clinical, biomedical, and translational science,'' \emph{Clinical and translational science}, vol.~15, no.~8, pp. 1848--1855, 2022.

\bibitem{Abuoda}
G.~Abuoda, S.~Thirumuruganathan, and A.~Aboulnaga, ``Accelerating entity lookups in knowledge graphs through embeddings,'' in \emph{IEEE 38th International Conference on Data Engineering (ICDE)}, vol.~6, no.~3, 2022, pp. 1111--1123.

\bibitem{hoyt2020extension}
H.~C. Tapley, M.~Christopher, V.~Nicole, D.~Domingo-Fern{\'a}ndez, H.~Matthew, and C.~Viswa, ``Extension of roles in the chebi ontology,'' \emph{ChemRxiv}, 2020.

\bibitem{dashplotly}
\BIBentryALTinterwordspacing
P.~T. Inc., ``Dash layout,'' [Online; accessed 17-May-2024]. [Online]. Available: \url{https://dash.plotly.com/layout}
\BIBentrySTDinterwordspacing

\bibitem{selkoe1994normal}
D.~J. Selkoe, ``Normal and abnormal biology of the! b-amyloid precursor protein.'' \emph{Annual review of neuroscience}, 1994.

\bibitem{querfurth2010alzheimer}
H.~W. Querfurth and F.~M. LaFerla, ``Alzheimer's disease,'' \emph{New England Journal of Medicine}, vol. 362, no.~4, pp. 329--344, 2010.

\bibitem{world2019risk}
{World Health Organization} \emph{et~al.}, \emph{Risk reduction of cognitive decline and dementia: WHO guidelines}.\hskip 1em plus 0.5em minus 0.4em\relax World Health Organization, 2019.

\bibitem{jack2010hypothetical}
J.~Clifford, K.~David, J.~William, S.~Leslie, A.~Paul, W.~Michael, P.~Ronald, and T.~John, ``Hypothetical model of dynamic biomarkers of the alzheimer's pathological cascade,'' \emph{The Lancet Neurology}, vol.~9, no.~1, pp. 119--128, 2010.

\bibitem{drugbank}
``Drugbank,'' \url{https://go.drugbank.com/drugs}, {Accessed Aug. 16, 2024.}

\bibitem{bitzinger2016species}
D.~I. Bitzinger, M.~Gruber, S.~T{\"u}mmler, B.~Michels, A.~Bundscherer, S.~Hopf, B.~Trabold, B.~M. Graf, and Y.~A. Zausig, ``Species-and concentration-dependent differences of acetyl-and butyrylcholinesterase sensitivity to physostigmine and neostigmine,'' \emph{Neuropharmacology}, vol. 109, pp. 1--6, 2016.

\bibitem{malamed2010sedation}
S.~F. Malamed \emph{et~al.}, \emph{Sedation: a guide to patient management}.\hskip 1em plus 0.5em minus 0.4em\relax Mosby Elsevier, 2010.

\bibitem{cacabelos2020pharmacogenetic}
R.~Cacabelos, ``Pharmacogenetic considerations when prescribing cholinesterase inhibitors for the treatment of alzheimer’s disease,'' \emph{Expert Opinion on Drug Metabolism \& Toxicology}, vol.~16, no.~8, pp. 673--701, 2020.

\bibitem{bansal2023hmg}
A.~B. Bansal and M.~Cassagnol, \emph{HMG-CoA Reductase Inhibitors}.\hskip 1em plus 0.5em minus 0.4em\relax Treasure Island, FL: StatPearls Publishing, 2024, [Updated 2023 Jul 3]. In: StatPearls [Internet].

\bibitem{cheng2020peripheral}
Y.~Cheng, D.~Tian, and Y.~Wang, ``Peripheral clearance of brain-derived a$\beta$ in alzheimer's disease: pathophysiology and therapeutic perspectives,'' \emph{Translational Neurodegeneration}, vol.~9, no.~1, pp. 1--14, 2020.

\bibitem{hampel2021amyloid}
H.~Hampel, J.~Hardy, K.~Blennow, C.~Chen, G.~Perry, S.~H. Kim, V.~L. Villemagne, P.~Aisen, M.~Vendruscolo, T.~Iwatsubo \emph{et~al.}, ``The amyloid-$\beta$ pathway in alzheimer’s disease,'' \emph{Molecular psychiatry}, vol.~26, no.~10, pp. 5481--5503, 2021.

\bibitem{salas2021quinacrine}
M.~Salas~Rojas, R.~Silva~Garcia, E.~Bini, V.~Perez de~la Cruz, J.~C. Leon~Contreras, R.~Hernandez~Pando, F.~Bastida~Gonzalez, E.~Davila-Gonzalez, M.~Orozco~Morales, A.~Gamboa~Dominguez \emph{et~al.}, ``Quinacrine, an antimalarial drug with strong activity inhibiting sars-cov-2 viral replication in vitro,'' \emph{Viruses}, vol.~13, no.~1, p. 121, 2021.

\bibitem{arumugham2023third}
V.~B. Arumugham, R.~Gujarathi, and M.~Cascella, \emph{Third-Generation Cephalosporins}.\hskip 1em plus 0.5em minus 0.4em\relax Treasure Island, FL: StatPearls Publishing, Jan. 2024, updated 2023 Jun 4.

\bibitem{biomedical2019toward}
{Biomedical Data Translator Consortium}, ``Toward a universal biomedical data translator,'' \emph{Clinical and Translational Science}, vol.~12, no.~2, p.~86, 2019.

\bibitem{NRRenci}
\BIBentryALTinterwordspacing
{TRANSLATOR RENCI}, ``Name resolver,'' [Online; accessed 17-May-2024]. [Online]. Available: \url{https://name-resolution-sri.renci.org/docs}
\BIBentrySTDinterwordspacing

\bibitem{tipney2010introduction}
H.~Tipney and L.~Hunter, ``An introduction to effective use of enrichment analysis software,'' \emph{Human genomics}, vol.~4, pp. 1--5, 2010.

\bibitem{UNIPROTKB}
\BIBentryALTinterwordspacing
{UniProtKB}, ``P81908,'' [Online; accessed 17-July-2024]. [Online]. Available: \url{https://www.uniprot.org/uniprotkb/P81908/entry2}
\BIBentrySTDinterwordspacing

\bibitem{colovic2013acetylcholinesterase}
M.~B. Colovic, D.~Z. Krstic, T.~D. Lazarevic-Pasti, A.~M. Bondzic, and V.~M. Vasic, ``Acetylcholinesterase inhibitors: pharmacology and toxicology,'' \emph{Current neuropharmacology}, vol.~11, no.~3, pp. 315--335, 2013.

\end{thebibliography}

\newpage
\appendix

\subsection{Additional UI Functionality } \label{sec:nametocurie}
\begin{figure}[http]
  \centering
  \fbox{\includegraphics[trim={0 14cm 0  1.3cm},clip,width=\linewidth]{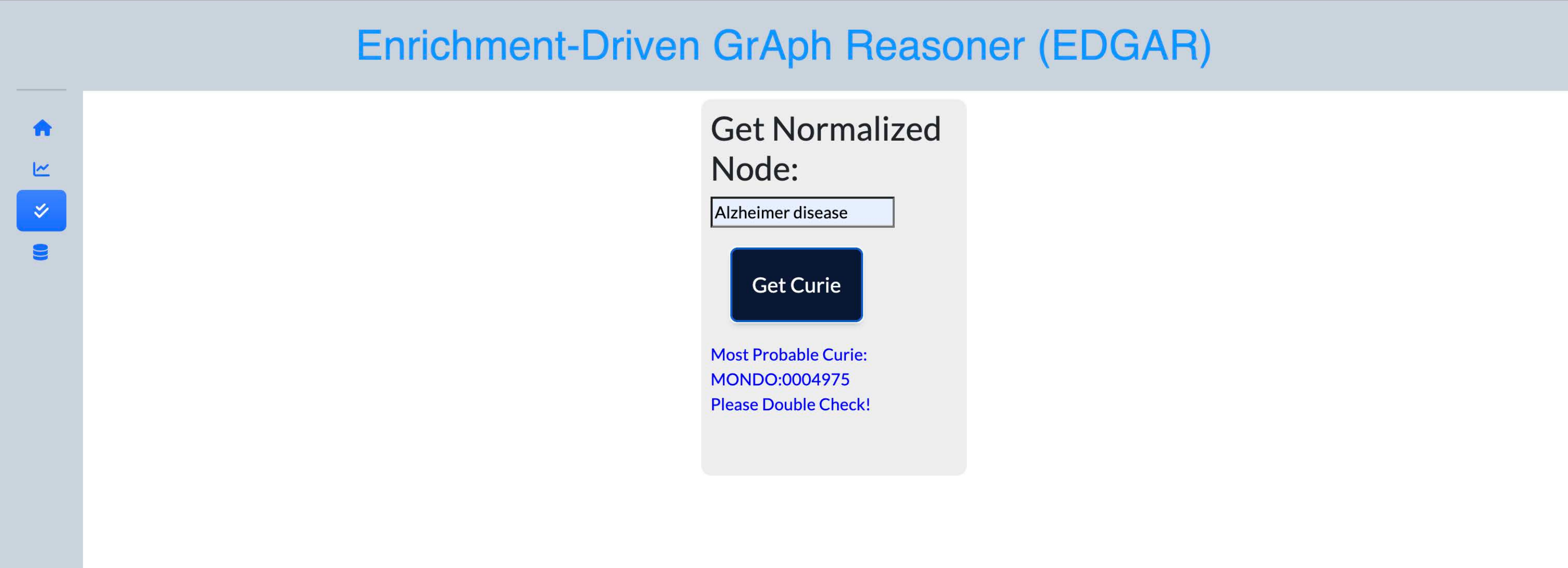}}
  \caption{UI finds TRAPI-compliant curies for biomedical terms.}
  \label{fig:ui_curie}
\end{figure}

The name-to-curie module in Figure~\ref{fig:ui_curie} constructs Biolink-compliant identifiers of nodes for a lexical input. It accepts biomedical terms and returns the top matching curie using the NCATS Biomedical Data Translator \cite{biomedical2019toward} Name Resolver API \cite{NRRenci}.  This module was used to resolve ``Alzheimer'' to ``MONDO:00004975".

\subsection{Scalability Considerations } \label{sec:scalability}
Given the scale of the ROBOKOP graph, we apply some constraints to the EDGAR implementation.

\paragraph{\textbf{Graph database}} \label{para:redundant_graph}
We wrapped up the EDGAR algorithm in AnswerCoalesce (AC), a TRAPI tool and database designed to tag query insights by their commonalities and aid interpretation. 
The AC database is populated with a version of ROBOKOP KG \cite{Bizonetal19} optimized to incorporate preprocessed redundant edges. For every edge in ROBOKOP KG, additional edges were added for every permutation of logically inferrable representations of relationships. This was accomplished by using Biolink to determine ancestors of predicates, qualifiers, and qualifier values of existing edges. This technique allows EDGAR to return valid results for queries that do not match the edges of ROBOKOP KG exactly but could be inferred, without having to compute those relationships at runtime. As opposed to other databases, the AC database was processed pretty heavily to make these types of queries faster.

\paragraph{\textbf{Node constraints}} We define a set of nodes to exclude as possible answers to the enrichment analysis. 
These nodes include entities such as Homo sapiens (NCBITaxon:9606), chemical entities (CHEBI:24431), and broad broad molecular entities (CHEBI:33304).  Even if these nodes are returned as enriched, their application will not lead to useful results. 
In addition to these handcrafted node constraints, we added the flexibility for users to specify other node sets for inclusion or exclusion.
While down-weighting these nodes during enrichment could be a valid approach, the decision to exclude them entirely was made for practical reasons. Since entities like "Homo sapiens" or broad chemical categories are often overrepresented and provide minimal interpretive value, excluding them upfront simplifies the analysis, ensuring that meaningful and actionable results are highlighted. Additionally, by giving users the flexibility to define node sets for exclusion or inclusion, we ensure that the analysis is adaptable to different use cases, improving its relevance and precision.

\paragraph{\textbf{Edge constraints}} 
Similarly, we introduce the option of disallowing particular relationships for enrichment.  While it may be true, for instance, that many chemicals for a disease have similar side effects, the existence of that side effect is not mechanistically helpful in determining a new treatment.  For this reason, in our experiments, we filter the similar relationships \textit{contraindicated\_for}, \textit{causes}, \textit{biomarker\_for}, \textit{contributes\_to}, \textit{has\_adverse\_event}, and \textit{causes\_adverse\_event}. Additionally, we filter the superclass hierarchy of each edge constraint.

\paragraph{\textbf{Approximating hypergeometric distribution}} Due to the large size of the KG, we used the Poisson approximation of the hypergeometric distribution for computational efficiency. 
The Poisson approximation \( P(X = k) \approx \frac{ e^{-\lambda}\lambda^k}{k!} \) to the hypergeometric distribution holds when $N$ is large, $n,k \ll N$, and \( \lambda = \frac{nK}{N} \), hence
\begin{equation}
    \text{SF}(k-1) = 1 - \sum_{i=0}^{k-1} \frac{e^{-\lambda}\lambda^i}{i!}. 
\end{equation}

\paragraph{\textbf{P-value threshold ($p_0$)}} $p_0$ in enrichment analysis is crucial to controlling the false positive rate, ensuring that the identified associations are statistically significant and not due to random variation \cite{tipney2010introduction}. We set $p_0$ to be fairly small at $10^{-5}$. This practice focuses on robust and meaningful results, which is particularly important in our ROBOKOP large graph data, enhancing reliability across studies. Additionally, we threshold the overall number of enrichment-based rules to 1,000.  The p-value and rules cutoffs can be user-specified.

\subsection{Time Complexity}

The runtime for processing queries through the EDGAR UI is influenced by several stages in the pipeline, which include: 

\begin{itemize}
    \item \textbf{Lookup Stage}: The initial retrieval of records based on the query. This stage's index search has an average time complexity of \(O(1)\). Let us call the number of resulting nodes $n$.
    
    \item \textbf{Enrichment Stage}: After lookup, results undergo enrichment calculation and filtering. The complexity here could approach \(O(n)\).
    
    \item \textbf{Inference Stage}: Generates new insights from the one-hop neighbors of the enriched results that satisfy the node and edge constraints. As a result, the complexity for this stage is \(O(m)\), where $m$ depends on the number of commonalities available in the KG.
\end{itemize}

To illustrate, we ran four sample queries and recorded the number of results and time taken in each stage. The table \ref{table:runtime_analysis} below summarizes the findings for Alzheimers (MONDO:0004975), Asthma (MONDO:0004979), Reduced circulating 4-Hydroxyphenylpyruvate dioxygenase activity (HP:0003637) and multiple endocrine neoplasia type 2A (DOID:0050430):
\begin{table*}[th]
    \centering
    \caption{Runtime Analysis for Various Queries}
    \label{table:runtime_analysis}
    \renewcommand{\arraystretch}{0.5} 
    \begin{tabular}{lllll}
    \toprule
    \textbf{Query} & \textbf{Lookup ($n$)} & \textbf{Enrichment Pre/Post-Filtering  ($m$)} & \textbf{Inference} & \textbf{Runtime (secs)} \\
    \hline
     \(\textcolor{red}{\text{HP:0003637}} \xleftarrow{r} \textcolor[HTML]{3166FF}{\text{?Gene}}\) & 1 & 545/100 & 2 & 4.8 \\
    \(\textcolor{red}{\text{MONDO:0004975}} \xleftarrow{r}\textcolor[HTML]{3166FF}{\text{?Drug}}\) & 14 & 1081/21 & 1246 & 9 \\
    \(\textcolor{red}{\text{MONDO:0004979}} \xleftarrow{r} \textcolor[HTML]{3166FF}{\text{?Drug}}\) & 49 & 2371/116 & 1805 & 20 \\
    \(\textcolor{red}{\text{DOID:0050430}} \xleftarrow{r} \textcolor[HTML]{3166FF}{\text{?Gene}}\) & 15 & 25029/97 & 2945 & 80 \\
    \bottomrule
    \end{tabular}
\end{table*}

This suggests that EDGAR does not scale with the overall graph but rather with the number of triples at each stage because of the index search.

\subsection{Intermediate Results from EDGAR}
\begin{table}[tp]
\centering
\footnotesize
    \caption{Lookup (Sec. \ref{sec:lookup_one}) returns 14 drugs connected to Alzheimer's Disease via a `treats' edge.}
    \label{tab:lookup-table}
    \begin{tabular}{@{}llll@{}}
    \toprule
   \multicolumn{2}{c}{\textbf{Drug (name and curie)}}& \textbf{Knowledge source} &  \\ \midrule
    {\color[HTML]{3166FF} O-acetyl-L-carnitine} & {\color[HTML]{3166FF} CHEBI:57589} & {drugcentral} \\ \addlinespace[1mm]
     {\color[HTML]{3166FF} rivastigmine} & {\color[HTML]{3166FF} CHEBI:8874} & {drugcentral} \\ \addlinespace[1mm]
   {\color[HTML]{3166FF} (-)-selegiline} & {\color[HTML]{3166FF} CHEBI:9086} & {hetionet} \\ \addlinespace[1mm]
    {\color[HTML]{3166FF} benzatropine} & {\color[HTML]{3166FF} CHEBI:3048} & {hetionet} \\ \addlinespace[1mm]
   {\color[HTML]{3166FF} acetylcholine} & {\color[HTML]{3166FF} CHEBI:15355} & {drugmechdb} \\ \addlinespace[1mm]
     {\color[HTML]{3166FF} ropinirole} & {\color[HTML]{3166FF} CHEBI:8888} & {hetionet} \\ \addlinespace[1mm]
     {\color[HTML]{3166FF} memantine} & {\color[HTML]{3166FF} CHEBI:64312} & {hetionet} \\ \addlinespace[1mm]
    {\color[HTML]{3166FF} quetiapine} & {\color[HTML]{3166FF} CHEBI:8707} & {hetionet} \\ \addlinespace[1mm]
   {\color[HTML]{3166FF} LECANEMAB} & {\color[HTML]{3166FF} UNII:12PYH0FTU9} & {drugcentral} \\ \addlinespace[1mm]
   {\color[HTML]{3166FF} tacrine} & {\color[HTML]{3166FF} CHEBI:45980} & {drugcentral} \\ \addlinespace[1mm]
    {\color[HTML]{3166FF} haloperidol} & {\color[HTML]{3166FF} CHEBI:5613} & {hetionet} \\ \addlinespace[1mm]
    {\color[HTML]{3166FF} ADUCANUMAB} & {\color[HTML]{3166FF} UNII:105J35OE21} & {drugcentral} \\ \addlinespace[1mm]
   {\color[HTML]{3166FF} galanthamine} & {\color[HTML]{3166FF} CHEBI:42944} & {drugcentral} \\ \addlinespace[1mm]
 {\color[HTML]{3166FF} donepezil} & {\color[HTML]{3166FF} CHEBI:53289} & {hetionet} \\ \bottomrule
    \end{tabular}
\end{table}

\begin{table}[htp]
\centering
    \footnotesize 
    \caption{Node property enrichment found 8 enriched ChEBI role properties.}
    \label{tab:enriched-properties-table}
    \setlength{\tabcolsep}{1.5pt}
    \begin{tabular}{@{}lll@{}}
    \toprule
    \textbf{Subset of lookup drugs} & \textbf{Enriched node property} & \textbf{P-value} \\ \midrule
    \textcolor[HTML]{3166FF}{3048, 64312, 15355, 8707,} & neurotransmitter agent & 1.63e-9 \\
   \textcolor[HTML]{3166FF}{5613, 8888, 8874, 42944} &&\\\addlinespace[1.5mm]
\textcolor[HTML]{3166FF}{64312, 8888, 5613, 3048} & antidyskinesia agent & 1.03e-8 \\ \addlinespace[1.5mm]
\textcolor[HTML]{3166FF}{64312, 8888, 8707, 5613} &dopaminergic agent & 7.64e-7  \\ \addlinespace[1.5mm]
    \textcolor[HTML]{3166FF}{64312, 8888, 3048} & antiparkinson drug & 1.87e-6 \\ \addlinespace[1.5mm]
  \textcolor[HTML]{3166FF}{15355, 8874, 3048, 42944}&  cholinergic drug & 2.33e-6 \\ \addlinespace[1.5mm]
   \textcolor[HTML]{3166FF}{3048, 64312, 8707, 5613, 8888} & central nervous system drug & 2.58e-6 \\ \addlinespace[1.5mm]
   \textcolor[HTML]{3166FF}{53289, 45980, 8874, 42944}&  EC 3.1.1 carboxylic ester & \\
    & hydrolase inhibitor & 6.89e-6 \\ \addlinespace[1.5mm]
  \textcolor[HTML]{3166FF}{53289, 8874, 42944} &  EC 3.1.1.8 cholinesterase inhibitor & 8.98e-6\\ \bottomrule
    \end{tabular}
\end{table}

\begin{table*}[thp]
\centering
    \footnotesize
    \caption{Graph enrichment found 12 enriched nodes. First column lists the relevant subset of ChEBI drugs from Table \ref{tab:lookup-table}.}
    \label{tab:enrich-table}
    \begin{tabular}{@{}llllll@{}}
    \toprule
    \textbf{Subset of lookup drugs} & \textbf{Relationship (Edge)} & \multicolumn{2}{c}{\textbf{Enriched node (Name and curie)}}  & \textbf{P-value}  \\  \midrule
   \textcolor[HTML]{3166FF}{8874, 53289, 42944, 45980}  & Decreases activity of & BCHE & NCBIGene:1,00033901 &2.68e-8 \\
\addlinespace[1.5mm]
   \textcolor[HTML]{3166FF}{8874, 53289, 42944, 45980}  & Binds to & CHLE\_HORSE &  UniProtKB:P81908 & 6.12e-7 & \\
  & & Cholinesterase (sprot)&& \\ \addlinespace[1.5mm]
   \textcolor[HTML]{3166FF}{8874, 64312, 3048, 8888, 9086}   & Ameliorates condition of & Parkinson disease & MONDO:0005180 & 6.32e-7 \\ \addlinespace[1.5mm]
   \textcolor[HTML]{3166FF}{8874, 53289, 42944} & Treats & Lewy body dementia & MONDO:0007488 & 1.94e-6 \\ \addlinespace[1.5mm]
 \textcolor[HTML]{3166FF}{8874, 53289, 42944, 45980} & Decreases activity of & Ache &   NCBIGene:11423 & 2.08e-6  \\  \addlinespace[1.5mm]
  \textcolor[HTML]{3166FF}{8874, 53289, 42944, 45980}  & Binds to & ACHE & NCBIGene:43 & 2.08e-6  \\ \addlinespace[1.5mm]
 \textcolor[HTML]{3166FF}{8874, 53289, 42944, 45980} &  Decreases activity of & ACES\_TETCF & UniProtKB:P04058 &  2.08e-6 \\ 
   &&Acetylcholinesterase (sprot) \\ \addlinespace[1.5mm]
   \textcolor[HTML]{3166FF}{8874, 15355, 53289, 45980, 42944} & Affects abundance of & BCHE &  NCBIGene:590 & 4.78e-6 \\ \addlinespace[1.5mm]
   \textcolor[HTML]{3166FF}{8874, 53289, 42944, 45980} &Decreases activity of&Ache &  NCBIGene:83817 &  6.42e-6  \\ \addlinespace[1.5mm]
   \textcolor[HTML]{3166FF}{53289, 42944, 45980} & Decreases activity of &CHRNE &  NCBIGene:1145 & 6.49e-6  \\  \addlinespace[1.5mm]
   \textcolor[HTML]{3166FF}{15355, 3048, 45980, 8707, 5613}& Related to &Chrm1 &  NCBIGene:25229 &  7.46e-6  \\\addlinespace[1.5mm]
\textcolor[HTML]{3166FF}{8707, 15355, 5613, 45980} & Related to & Chrm4 &   NCBIGene:25111 & 9.31e-6 \\ \bottomrule
    \end{tabular}%
\end{table*}

Recall that EDGAR has three components. Table~\ref{tab:lookup-table} lists the output of the first stage, the lookup stage, which identifies drugs with therapeutic potential for AD within a one-hop distance. This yielded a total of 14 drug candidates, each linked to AD through a `treats' edge. For easier reading, we consistently color these 14 drugs as {\color[HTML]{3166FF} blue} whenever they re-appear in other tables. 
We then consider both types of enrichment analysis.
The \textbf{graph enrichment} analysis found 13 enriched nodes, listed in Table \ref{tab:enrich-table}. Two of the biomedical entities are Disease entities and 11 are distinct Gene/GeneProduct/ProteinCoding. 
The top 3 Gene/GeneProduct/ProteinCoding are  Butyrylcholinesterase-human (BCHE), Butyrylcholinesterase-Rattus (Bche), Cholinesterase-horse (CHLE), in which the 3 cholinesterase enables acetylcholinesterase activity and generally degrade neurotoxic organophosphate esters \cite{UNIPROTKB}. Acetylcholine is a primary neurotransmitter involved in learning, thinking, judgment, and memory while Cholinesterase inhibitors are used to increase acetylcholine levels in the brain by preventing its breakdown \cite{colovic2013acetylcholinesterase}.

On the other hand, the \textbf{node property enrichment}  produced 8 differentially expressed ChEBI roles in Table~\ref{tab:enriched-properties-table}. The top ChEBI role, neurotransmitter agents, was expressed in 8 (benzatropine, memantine, acetylcholine, quetiapine, haloperidol, ropinirole, rivastigmine, and galanthamine) out of the 14 treatment drugs in Table~\ref{tab:lookup-table}. Neurotransmitter agents are substances that act as agonists, antagonists, degradation inhibitors, uptake inhibitors, depleters, precursors, and modulators of receptor function to influence neurotransmitter activity \cite{drugbank}.

The top enrichment results in both cases are neurotransmitter agents which are directly related to the pathophysiology and treatment of AD. Understanding and targeting neurotransmitter systems can help manage symptoms and improve the quality of life for individuals with AD. Hence, we postulate that neurotransmitter agents in the lookup drugs are important pathways in AD. 



\end{document}